# Observation of a mesoscopic magnetic modulation in chiral $Mn_{1/3}NbS_2$

Sunil K. Karna[1], F. N. Womack[1], R. Chapai[1], D. P. Young[1], M. Marshall[2], Weiwei Xie[2], D. Graf[3], Yan Wu[4], Huibo Cao[4], L. DeBeer-Schmitt[4], P. W. Adams[1], R. Jin[1], and J. F. DiTusa[1]

[1]Department of Physics and Astronomy, Louisiana State University, Baton Rouge, Louisiana 70803, USA
[2]Department of Chemistry, Louisiana State University, Baton Rouge, Louisiana 70803, USA
[3]National High Magnetic Field Laboratory, Florida State University, Tallahassee, Florida 32310, USA
[4]Neutron Scattering Division, Oak Ridge National Laboratory, Oak Ridge, TN 37831, USA

Abstract

We have investigated the structural, magnetic, thermodynamic, and charge transport properties of $Mn_{1/3}NbS_2$ single crystals through x-ray and neutron diffraction, magnetization, specific heat, magnetoresistance, and Hall effect measurements. $Mn_{1/3}NbS_2$ displays a magnetic transition at $T_C \sim 45$ K with highly anisotropic behavior expected for a hexagonal structured material. Below $T_C$, neutron diffraction reveals increased scattering near the structural Bragg peaks having a wider $Q$-dependence along the $c$-axis than the nuclear Bragg peaks. This indicates helimagnetism with a long pitch length of ~250 nm (or a wavevector $q \sim 0.0025$ Å$^{-1}$) along the $c$-axis. This $q$ is substantially smaller than that found for the helimagnetic state in isostructural $Cr_{1/3}NbS_2$ (0.015 Å$^{-1}$). Specific heat capacity measurements confirm a second order magnetic phase transition with a substantial magnetic contribution that persists to low temperature. The large low-temperature specific heat capacity is consistent with a large density of low-lying magnetic excitations that are likely associated with topologically interesting magnetic modes. Changes to the magnetoresistance, the magnetization, and the magnetic neutron diffraction, which become more apparent below 20 K, imply a modification in the character of the magnetic ordering corresponding to the magnetic contribution to the specific heat capacity. These observations signify a more complex magnetic structure both at zero and finite fields for $Mn_{1/3}NbS_2$ than for the well-investigated $Cr_{1/3}NbS_2$.

## I. Introduction

Chiral structured magnetic materials have generated considerable attention due to the topological nature of their magnetic structures, which include the formation of noncollinear and noncoplanar spin textures with long length-scale modulations [1-5]. These magnets have promising ingredients for future nanometer scale quantum-information technology applications. Here, the chirality of the magnetism is generated by the crystal symmetry and the related spin-orbit coupling that create an antisymmetric exchange interaction known as the Dzyaloshinskii-Moriya (DM) interaction. The DM interaction is typically one or two orders of magnitude weaker than the isotropic exchange coupling [6, 7], and the competition between the DM interaction strength, $D$, and the exchange interaction, $J$, causes the appearance of spin textures with a left- or right-handed chirality, depending on the sign of $D$ and the handedness of the crystal structure [8]. When an external magnetic field is applied above a small threshold value, the helimagnetic (HM) ordering can be modulated into particle-like spin textures, such as a skyrmion or magnetic soliton lattice [2, 9].

The most well-known case of topological magnetism occurs in the cubic structured, *B20*, silicides and germanides [2-4], where a skyrmion lattice state can be accessed for small fields and temperatures near the Curie temperature, $T_C$. More recently, topology was discovered to be key to the magnetic structures found in hexagonal and chiral structured $Cr_{1/3}NbS_2$, whose strong anisotropy instead favors a novel magnetic chiral magnetic soliton lattice phase when exposed to small fields [9, 10]. This unusual magnetic structure consists of a superlattice based on a periodic helical spin texture [9, 11, 12]. $Cr_{1/3}NbS_2$ is synthesized by intercalating Cr between the layers of planar 2H-type $NbS_2$, creating a crystal structure with the non-centrosymmetric and chiral space group $P6_322$ [13-15]. Its unique magnetic properties arise from the strong uniaxial anisotropy (easy plane) along with $S = 3/2$ $Cr^{3+}$ magnetic moments. In addition, there is a small density of electronic charge carriers creating a highly anisotropic metal. As a result, $Cr_{1/3}NbS_2$ displays a helical magnetic ground state below 127 K with a small wavevector, $q \sim 0.015$ Å$^{-1}$, along the *c*-axis [16, 17]. The application of small magnetic fields can lead to a simple conical magnetic phase for H // *c*, or a novel chiral soliton lattice (CSL) for H // *ab* [9].

The ability to intercalate a wide variety of elements between the van der Waals bonded layers of $NbS_2$ allows us to question if unique magnetic structures and behaviors can be accessed by intercalating other transition metal, TM, species, or if the same magnetic states are accessed as in $Cr_{1/3}NbS_2$. The wide variations to the helical and skyrmion lattice states found in the binary magnetic *B20's* leads us to posit that there will be fundamental modifications to the magnetic structures found in $TM_{1/3}NbS_2$ with changes in the intercalated species. Here, we report on the synthesis of Mn-intercalated $NbS_2$, $Mn_{1/3}NbS_2$, in single crystal form and the investigation of its magnetic, thermodynamic, and charge transport properties [14, 18-22]. The data presented here largely reinforce our supposition. We find a highly anisotropic magnetization and a magnetic transition at $T_C$ = 45 K evident in the magnetic susceptibility, specific heat, and resistivity. Neutron diffraction indicates a modulated magnetic ordering along the *c* axis, which is likely a long period helix with $q$~0.0025 Å$^{-1}$. Changes to the magnetization, charge transport, and specific heat capacity with cooling below $T_C$ and with the application of magnetic fields reveal behavior that is more complex than was observed in isostructural $Cr_{1/3}NbS_2$, suggesting a significant difference in the magnetic states both at zero and finite field.

## II. Experimental details

Single crystals of $Mn_{1/3}NbS_2$ were grown by iodine vapor transport with a central furnace temperature of 950 °C and a tube end temperature of 800 °C, as described in Ref. [18]. Powder X-ray diffraction (PXRD) measurements on crushed crystals were carried out on a PANalytical Empyrean multi-stage X-ray diffractometer with Cu K$\alpha$ radiation ($\lambda$ =1.54059 Å). These confirmed the $Nb_3MnS_6$ hexagonal crystal structure (space group $P6_322$), as shown in Fig. 1a and c with no apparent impurity phases. Single crystal x-ray diffraction was performed on crystals mounted on the tip of a Kapton loop. Room temperature (296 K) data were collected on a Bruker Apex II X-ray diffractometer with Mo K$\alpha_1$ radiation ($\lambda$=0.71073 Å). Data were collected over a full sphere of reciprocal space with 0.5° scans in ω and an exposure time of 10 s per frame using SMART software for data acquisition. The 2θ range extended from 4° to 75°. Intensities were extracted and corrected for Lorentz and polarization effects with the SAINT program. Numerical absorption corrections were accomplished with XPREP based on face-indexed absorption [23]. The crystal structures were solved using the SHELXTL package with direct methods and refined by full-matrix least-squares on F$^2$ [24]. Electron density Fourier maps were generated by WINGX [25]. The single crystal X-ray diffraction confirmed the crystal structure determined from PXRD. After locating all the atomic positions, the displacement parameters were refined as anisotropic, and weighting schemes were applied during the final stages of the refinement. The refinement showed that there was some site disorder apparent for the intercalated Mn. The refinement was consistent with 85% of the Mn on the expected $2c$ site (1/3, 2/3, 1/4) (as displayed in Fig. 1a), while an anomalous high electron density residual (+6.18 $e^-/\text{Å}^3$) on the $2b$ site (0, 0, 1/4) indicated partial occupancy of the Mn atoms. The single crystal refinement details are included in the Supplemental Material Tables S1-S3 [26]. To confirm the atomic distributions in $Mn_{1/3}NbS_2$, a Fourier map of electron density residuals is presented in Fig 1b, showing the difference between the observed ($F_{obs}$) and calculated ($F_{cal}$). The electron density residuals on both the $2c$ and $2b$ sites demonstrate that Mn atoms partially occupy both sites. To check that the ordered structure corresponds to a superlattice, we carefully examined the diffraction precession image of the (HK0) plane shown in Fig. 1d. The reflections marked with blue circles represent the trigonal $CdI_2$-type (1T-type) structure and the reflections indicated by yellow circles provide proof of the ordered superlattice structure type $Mn_{1/3}NbS_2$ with the $P6_322$ space group.

Chemical analyses were performed using a JEOL JSX-8230 SuperProbe electron probe microanalyzer (EPMA). This instrument allows simultaneous measurement via wavelength dispersive spectroscopy (WDS) and energy dispersive spectroscopy (EDS) techniques. WDS determined a chemical composition of $Nb_{2.97}Mn_{1.03}S_{6.03}$, hereafter referred to as $Mn_{1/3}NbS_2$. We carefully checked for any indication of iodine in the WDS measurements of these crystals, finding no signal above background. The magnetic structure of our single crystals was investigated via neutron diffraction performed at the four-circle diffractometer HB3A at Oak Ridge National Laboratory (ORNL). Reflections were collected at $T = 5$ K, 55 K, and 100 K, using a wavelength of 1.550 Å from a perfect bent Si-220 monochromator [27] for the magnetic structure determination. Refinements of the magnetic structure were performed with the FULLPROF suite [28]. In addition, we performed L-scans (in the usual notation (HKL) to identify the reciprocal lattice) in proximity to the (011) nuclear Bragg peak with a larger wavelength, $\lambda=2.541$Å (Si-111 monochromator), for a series of temperatures between 5 and 55 K.

Magnetization and magnetic susceptibility measurements with a magnetic field applied parallel and perpendicular to the crystallographic $c$-axis were carried out in a Quantum Design (QD) Magnetic Property Measurement System (MPMS) superconducting quantum interference device (SQUID) magnetometer. These included $ac$ susceptibility measurements performed at a frequency of 100 Hz with an $ac$ driving amplitude of 3.9 Oe. The electrical resistivity and magnetoreistance (MR) were measured on single crystals with contacts formed via conductive epoxy (Epotek H20E) and thin platinum wire. These measurements were standard four-terminal *ac* resistance measurements with a current of 3 mA at a frequency of 17 Hz applied parallel to the *ab* plane of our crystals. Data was collected in a QD Physical Property Measurement System (PPMS) in a 90 kOe superconducting magnet, and in a 350 kOe resistive magnet at the National High Magnetic Field Lab (NHMFL) in Tallahassee, FL. The MR was corrected for possible misalignment of the contacts by symmetrizing the data for positive and negative fields. Similarly, the Hall effect was measured with current and voltage contacts defined on the top of a *c*-axis oriented crystal. These measurements were carried out using an *ac* current of 3 mA and were corrected for misalignment of the leads by symmetrizing the data for positive and negative fields oriented along the crystallographic *c*-axis. The specific heat capacity was measured using a time-relaxation method in a QD PPMS between 2 and 100 K with magnetic fields of up to 90 kOe applied parallel and perpendicular to the *c*-axis.

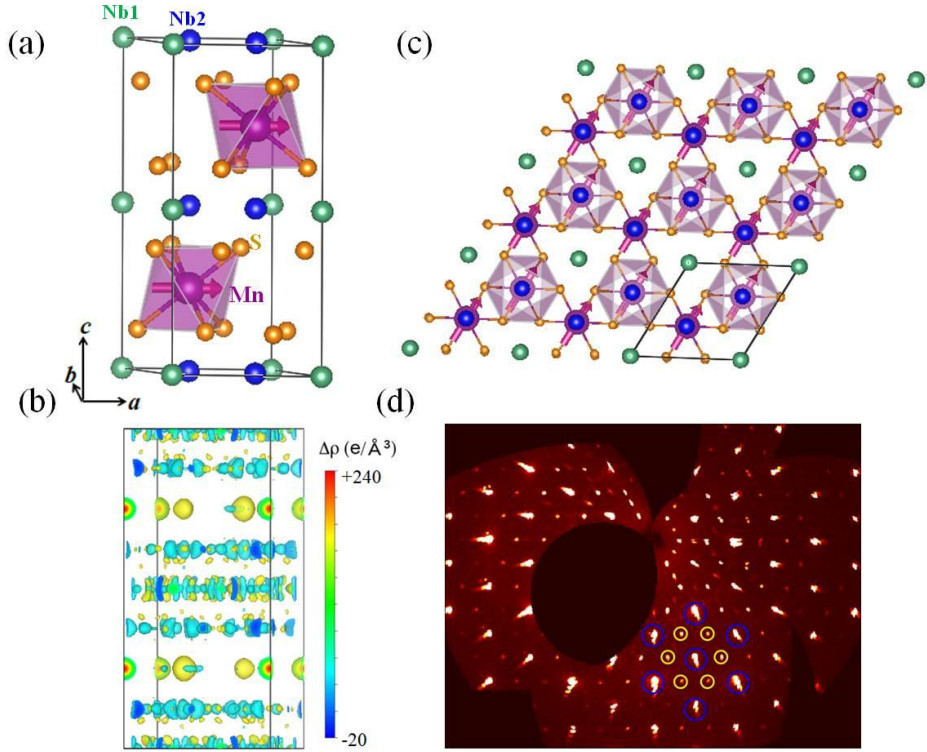

Fig. 1. Crystal and magnetic structure of Mn$_{1/3}$NbS$_2$. (a) Crystal structure demonstrating the intercalated Mn atoms which occupy the octahedral interstitial holes (*2c* site) between trigonal prismatic layers of 2H-NbS$_2$ in the ideal case. Arrows represent the Mn magnetic moments, as determined by a refinement of the neutron scattering data. (b) Fourier map of electron density residual [($F_{obs}$)-($F_{cal}$)] with a resolution of 0.05 e$^-$/Å$^3$. (c) Two dimensional view of the crystal structure along the *c*-axis highlighting the triangular network of Mn. The top Mn layer is indicated via shading of the surrounding octahedron of sulfur. (d) Diffraction precession image of the (HK0) plane of the reciprocal lattice. All of the resolved intensity peaks have been identified with the crystal lattice structure of the chiral space group *P6$_3$22*. Blue circles identify reflections expected for the CdI$_2$ structure type of the underlying NbS$_2$ lattice, while reflections identified with yellow circles confirm the symmetry expected for the *P6$_3$22* space group. The flaring associated with reflections such as those enclosed by blue circles is instrumental in origin.

**III. Results and discussions**

After establishing the crystal structure and the site occupancy for the intercalated Mn, the temperature dependence of the magnetic susceptibility, $\chi$, for *H* parallel and perpendicular to the *c*-axis was measured and is presented in Fig. 2(a), (c), and (d). The *T* dependence of the magnetic susceptibility indicates a magnetic transition at $T_C \sim 45$ K, and a large anisotropy below $T_C$, both of which are consistent with earlier measurements [14]. A fit of the Curie-Weiss (CW) law to the data for *H* =90 Oe parallel to the *c*-axis at temperatures between 150 and 350 K yields $\Theta \approx 61.6$ K [$\Theta \approx 59.7$ K for $H \perp c$-axis] as shown in the inset to Fig. 2a, indicating ferromagnetic interactions of the magnetic moments. The difference in $\Theta$ and $T_C$, with $\Theta > T_C$, likely reflects the layered structure of this material resulting in quasi-two-dimensional magnetic interactions. The best fit resulted in a Curie constant of $C = 2.88$ emu K/mol corresponding to a fluctuating magnetic moment of 4.80 $\mu_B$ [4.97 $\mu_B$ for $H \perp c$-axis], which suggests $Mn^{3+}$ with $S = 2$. The magnetic moment found from the CW fitting procedure is in good agreement with the 4.8-5.1 $\mu_B$ value reported in previous literature [18, 19], but is significantly larger than that predicted by electronic structure calculations (3.8 $\mu_B$) [29]. A very small anomaly at ~105 K, which corresponds to the magnetic transition temperature of $Mn_{1/4}NbS_2$, is observed in $\chi(T)$ and was avoided in the temperature range over which the fits were performed [30, 31]. The magnetic susceptibility was measured under both zero-field-cooled (ZFC) and field-cooled (FC) conditions at $H = 5$ Oe [Fig. 2 (c) and (d)] with a significant history dependence apparent below ~45 K as was reported in Ref. [20]. This is similar to what was observed in $Cr_{1/3}NbS_2$ [32]. The anisotropic nature of the low-temperature magnetism suggested by the layered crystal structure of $Mn_{1/3}NbS_2$ is also evident in $M(H)$, as demonstrated at 2 K in Fig. 2b. The anisotropy is similar to that observed in $Cr_{1/3}NbS_2$ [33]. The field dependence of $M$ resembles that of ferromagnetic materials with a saturation field near 0.5 kOe for $H$ perpendicular to the $c$-axis, and about 40 kOe for $H$ parallel to the $c$-axis. We note that the saturation fields we observe are substantially different from an earlier measurement [14], but are largely consistent more recently published data [20], and we are unsure of the reason for the discrepancy. The size of the saturated magnetic moment, ~4.1 $\mu_B$, is smaller than the magnetic moment determined from the Curie constant and is closer to expectations for a $Mn^{4+}$ ion, although it is somewhat larger than that expected from electronic structure calculations [29]. The discrepancy in the magnetic moments determined from $\chi(T)$ and the saturated magnetic moment suggests that the magnetism may have some itinerant character and we point out that a similar discrepancy has been reported in $Cr_{1/3}NbS_2$ [13, 19, 33].

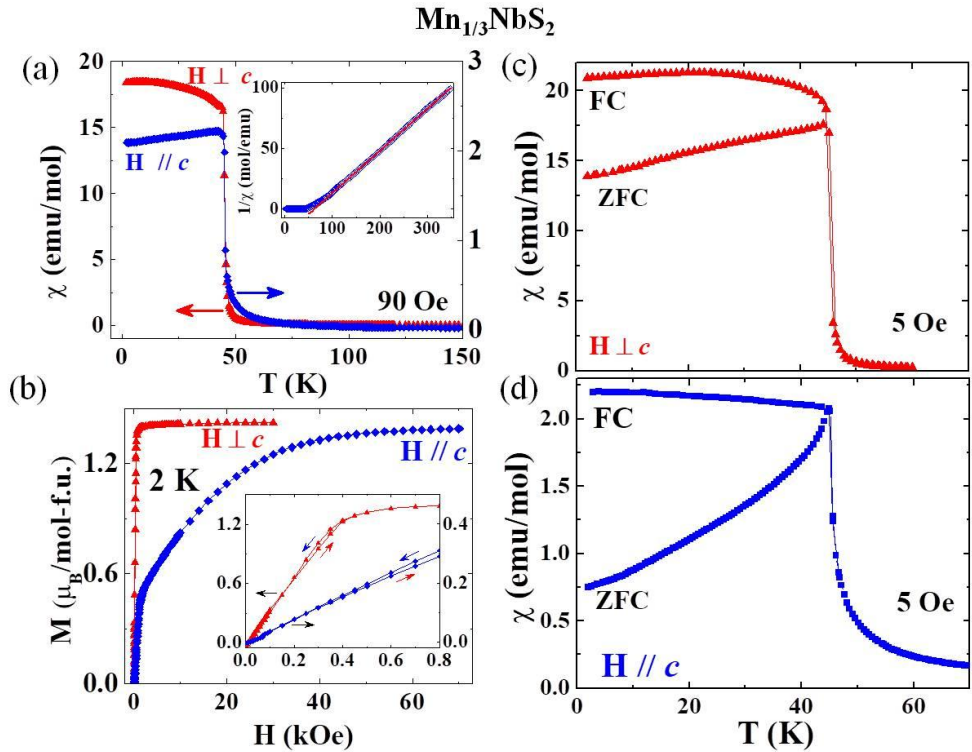

Fig. 2. Magnetic Susceptibility and Magnetization of $Mn_{1/3}NbS_2$. (a) Temperature, $T$, dependence and anisotropy of the magnetic susceptibility, $\chi$, taken with a field $H$=90 Oe. Inset: $1/\chi$ vs. $T$ for $H$ parallel to the $c$-axis. Line is a fit of the Curie Weiss form to the data for 150 <$T$<350 K. (b) Magnetization, $M$, vs. magnetic field, $H$, at 2 K with $H$ parallel and perpendicular to the $c$-axis. Inset: low field $M$. $T$ dependence of the field cooled, FC, and zero field cooled, ZFC, magnetic susceptibility at $H$ = 5 Oe with (c) $H \perp c$ and (d) $H // c$.

In light of the interesting magnetic structures found in $Cr_{1/3}NbS_2$, we have investigated the magnetic structure of $Mn_{1/3}NbS_2$ through single-crystal neutron diffraction. We collected 110 reflections at 5 K to characterize the magnetic ordering. In addition, we selected a weak nuclear Bragg peak (011) to track the magnetic order by scanning through (011) along the reciprocal $L$-direction in consideration of the overlap of the magnetic scattering and the nuclear reflections. Representative data are shown in Fig. 3(a), where a scan along the $L$-direction is displayed at 5 and 50 K. The scattering associated with the magnetic ordering is clearly observed at 5 K as the intensity has increased dramatically from that at 50 K. The increased scattering in proximity to a nuclear Bragg peak indicates that the magnetism is nearly ferromagnetic. However, a closer inspection of the data in Fig. 3(a) including fitting the data to a simple Gaussian reveals changes that are not consistent with a simple ferromagnetic ordering. The results of this straightforward analysis are presented in Fig. 3(b) where the full-width at half-maximum (FWHM) along $L$ and the integrated intensity, $I$, from the peak centered at (011) is displayed for temperatures between 5 and 55 K. Here, it is clear that $I(T)$ is consistent with a magnetic ordering at $T_C \sim 45$ K while the FWHM displays non-monotonic temperature dependence below $T_C$. The FWHM increases by as much as 50%, indicating a more complex magnetic ordering. Furthermore, we have also performed a magnetic and nuclear structure refinement based on 110 reflections collected at 5 K. Here we assume a ferromagnetic structure by ignoring the broadened magnetic scattering and include the fact that 85% of the Mn resides on the *2c* site. The refinement indicates that the magnetic moments are confined to the crystallographic *ab*-plane, similar to the case of $Cr_{1/3}NbS_2$ [10], and that the magnetic moment has a magnitude of 4.3(2) $\mu_B$, slightly higher than the magnetic moment obtained from NMR measurements [21]. A refinement to a long pitch length helical state with $q$ held constant at $(00q_c)$ with $q_c=0.0025$ Å$^{-1}$ resulted in a similar value for the magnetic moment that was also confined to the *ab* plane. This magnetic moment is similar to that found from the saturated magnetization.

The increased FWHM of the scattering evident below 50 K in Figs. 3(a) and (b) signifies a magnetic ordering that is inconsistent with long-range ferromagnetic order, which would result in magnetic scattering having the same FWHM as the nuclear Bragg peak. Because of the non-centrosymmetric crystal structure of $Mn_{1/3}NbS_2$, a significant DM interaction is anticipated, creating expectations for a helical magnetic state as in other chiral-structured magnetic materials. The increased FWHM along the *L*-direction (i.e., along the *c*-axis) implies helimagnetic ordering with a small wavevector, $\boldsymbol{q} = (00q)$, and our refinement indicates magnetic moments lying perpendicular to $\boldsymbol{q}$ (in the *ab*- plane). By fitting the magnetic contribution to the scattering in Fig. 3(a) to two Gaussians with a FWHM kept constant at the value determined by the width of the scattering peak at 50 K, we have made a rough estimate of $q \sim 0.0025$ Å$^{-1}$ corresponding to a helimagnetic pitch length of $\lambda \sim 250$ nm. We note that this $q$ is substantially smaller than that found in $Cr_{1/3}NbS_2$ (0.015 Å$^{-1}$) [10] and that our data are consistent with the $q$ (pitch length of $\lambda \sim 300$ nm) reported in Ref. [22] for polycrystalline $Mn_{1/3}NbS_2$ samples. In addition, the long period of the helimagnetic structure observed in $Mn_{1/3}NbS_2$ is compatible with expectations based upon electronic structure calculations [29]. We also note that there is a decrease in the FWHM of the (011) peak below 15 K, suggesting a possible change in the magnetic order at low temperatures.

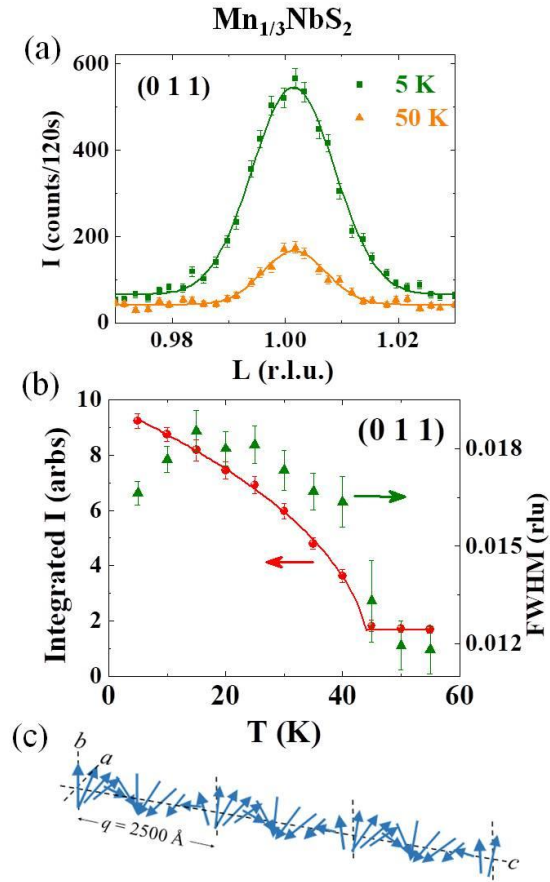

Fig. 3. Magnetic Structure of $Mn_{1/3}NbS_2$. (a) Neutron diffraction intensity, *I*, for a scan along the (0 1 *L*) direction in proximity to the (0 1 1) nuclear Bragg position. (b) *T*-dependence of integrated intensity and full-width at half-maximum (FWHM) of the (0 1 1) reflection indicating magnetic ordering with a critical temperature $T_C$ of 45 K. The solid line is a fit of a standard critical behavior model to the integrated intensity (see text). (c) Schematic illustration of the magnetic structure demonstrating the helical pitch along the *c*-axis.

The temperature dependence of the magnetic order parameter (Fig. 3(b)) can indicate the character of the magnetic ordering. Following a standard analysis, we fit the form $I = I_N + I_M(1 - T/T_C)^{2\beta}$ [34], where $I_N$ represents the temperature independent contribution from the nuclear reflection, and $I_M$ indicates the magnetic intensity at saturation, to the integrated intensity data of Fig. 3(b) for $T < T_C$. The solid line in Fig. 3(b) represents the best fit of this form to the data with a value for the critical exponent, $\beta = 0.23(3)$. This value of $\beta$ lies between that expected for the three-dimensional Heisenberg model ($\beta = 0.38$) and the two-dimensional Ising model ($\beta = 0.125$), and close to that expected from a two-dimensional XY model. This is a reasonable result for a layered material such as $Mn_{1/3}NbS_2$.

After characterizing the magnetic properties of our $Mn_{1/3}NbS_2$ crystals, we explored their charge transport properties to characterize the coupling between the magnetic and charge degrees of freedom. Fig. 4(a) shows the temperature dependence of the in-plane electrical resistivity, $\rho_{ab}$, at magnetic fields, $H$, between 0 and 90 kOe. Note that $\rho_{ab}$ exhibits metallic behavior with a residual resistivity ratio (RRR) of ~64 signaling high quality crystals with a long mean-free path of charge carriers. The behavior near the magnetic phase transition is similar to that of itinerant magnets with a sharp drop below $T_C$ suggesting a significant reduction in the magnetic fluctuation scattering with magnetic ordering. The application of field effectively smooths the anomaly around the critical temperature. The resistivity is reduced by as much as 40% for a 90 kOe and 70% for a 350 kOe [inset of Fig. 4(d)] field near $T_C$, which is a large negative magnetoresistance (MR) even when considering the expected reduction of scattering from magnetic fluctuations. The MR for in-plane currents is presented in Fig. 4(b), (c), and (d) for temperatures between 5 and 60 K for $H$ oriented parallel to the $c$-axis, as well as perpendicular to the $c$-axis and parallel to the current (longitudinal MR). The MR for these two orientations of $H$ are similar, as both are strongly negative at high fields. However, the MR for $H // c$ includes a positive *MR* contribution that is maximum at 10 K near 27 kOe [Fig. 4 (b) and (c)], suggesting that it may be due to scattering from magnetic moments that are canted along the *c*-axis with field. For comparison, we note that in isostructural $Cr_{1/3}NbS_2$, a positive MR has not been reported [35] for *H // c*, and the negative MR seen there for $H \perp c$ is approximately seven times smaller than we observe in $Mn_{1/3}NbS_2$ at 5 K and 70 kOe. The field dependence seen in Fig. 4 does not saturate despite *M(H)* saturating near 0.5 kOe [Fig. 2(b)], suggesting a mechanism more complex than a simple field reduction in spin-disorder scattering. We note that the MR for $H \perp c$ resembles a small power-law form, which is confirmed by fits of the data (Fig. 4(d)) that indicate a reduction of the power-law from 3/4 to 2/3 with increasing temperature from 5 to 40 K. A similar negative *MR* was observed in $Cr_{1/3}NbS_2$ for samples that are thinner than the helical pitch (≤ 47nm) [36]. However, we observe no hysteresis or acute jumps in the MR of our bulk single crystal of $Mn_{1/3}NbS_2$.

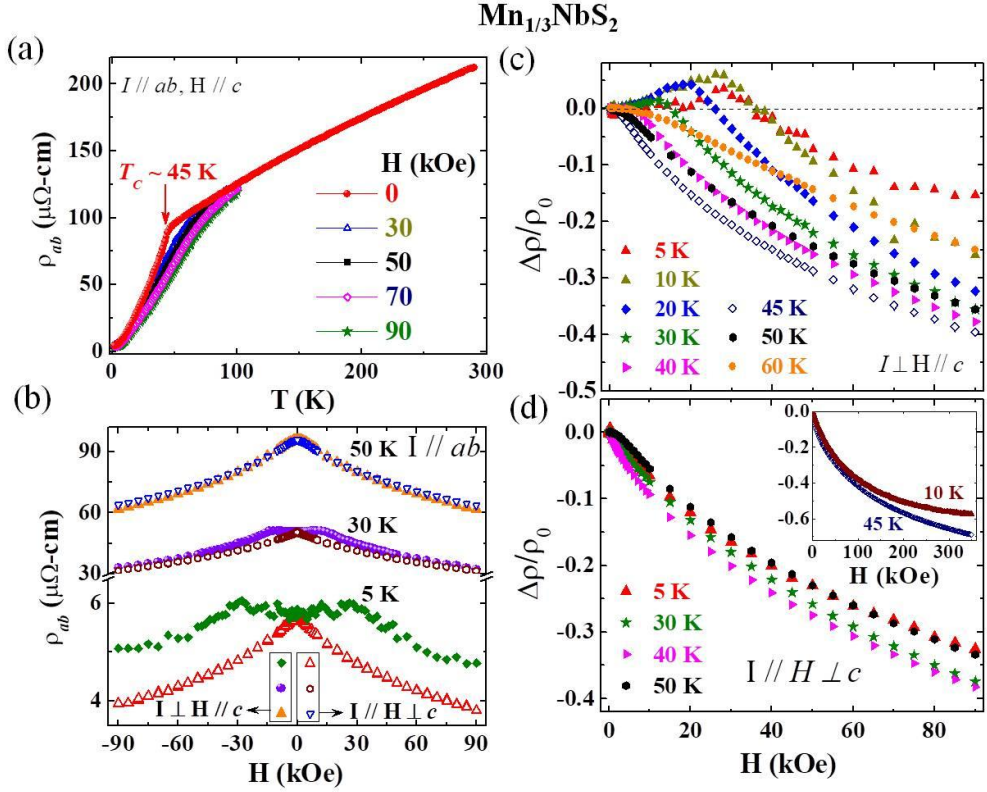

Fig. 4. Charge transport properties of $Mn_{1/3}NbS_2$. (a) Temperature dependence of the resistivity measured with current, $I$, in the $ab$-plane, $\rho_{ab}$ at magnetic fields applied parallel to the $c$-axis. (b) Magnetoresistance, $\rho_{ab}(H)$, with fields applied parallel (filled symbols) and perpendicular (open symbols) to $I$. Magnetoresistance, $\Delta\rho/\rho_0 = (\rho_{ab}(H) - \rho_{ab}(0)) / \rho_{ab}(0)$, (MR) measured with $H$ applied (c) parallel and (d) perpendicular to the $c$-axis. The solid lines are fits of the form $\Delta\rho/\rho \sim H^\alpha$ to the data. Inset: MR at 10 and 45 K for $H$ up to 350 kOe.

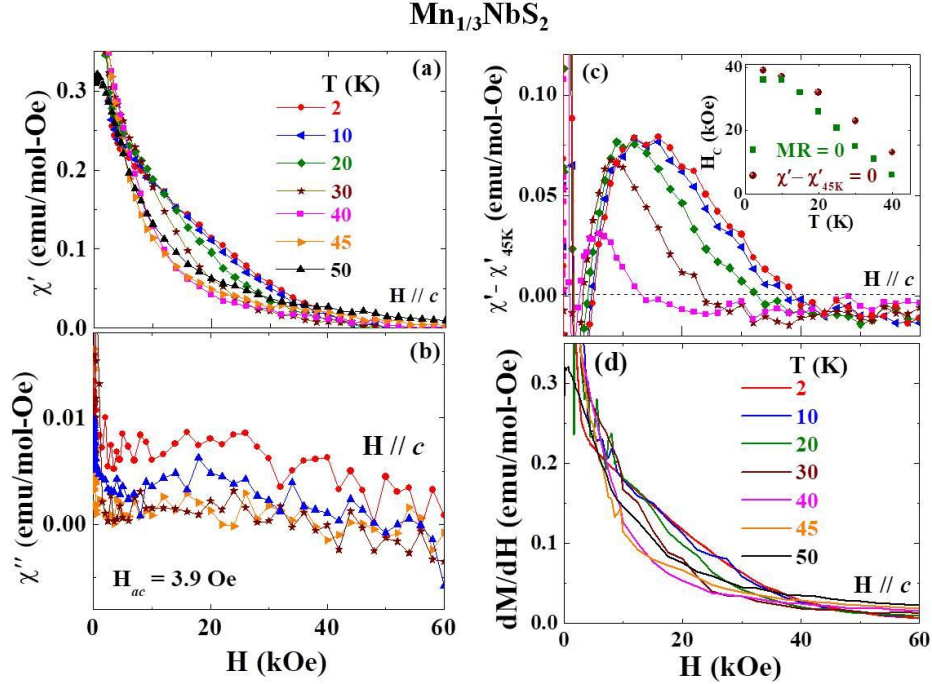

Fig. 5. Magnetic Field dependence of the *ac* magnetic susceptibility. (a) Real, χ′, and (b) imaginary χ″ parts of the *ac* magnetic susceptibility vs. magnetic field applied parallel to the *c*-axis. (c) χ′ after subtraction of χ′ at 45 K, χ′-χ′$_{45K}$. Inset: Zero crossing fields for the MR (see Fig. 4c) and χ′-χ′$_{45K}$ with *H* parallel to the *c*-axis vs. *T*. (d) Derivative of the magnetization with respect to *H*, *dM/dH* vs. *H* with *H* parallel to the *c*-axis.

To explore the origins of the positive MR for *H // c*, we have measured the *ac* susceptibility with the drive field in the same orientation over similar field and temperature ranges. Fig. 5(a) displays the field dependence of the real part, χ′, and Fig. 5(b) the imaginary part, χ″, of the *ac* magnetic susceptibility at the indicated temperatures. Here, a temperature dependent contribution to χ′(*H*) is observed between 10 and 40 kOe for *T*<*T$_C$* that is also apparent in the field derivative of the dc $M(H)$ in Fig. 5(d). To isolate this contribution, we present $\chi'(H)$ with the value at 45 K subtracted, $\chi'-\chi'_{45K}$ in Fig. 5(c). Presenting the data in this manner makes it clear that the additional contribution to $\chi'$ is similar to the positive contribution to the MR, increasing in magnitude and having an increased characteristic field with cooling. We have quantified this behavior in the inset to Fig. 5c which compares the field of the zero crossing in the MR $[(\rho-\rho_0)/\rho_0 = 0]$ to the zero crossing of $\chi'-\chi'_{45K}$. This comparison suggests a magnetic scattering origin for the positive component of the MR ($H // c$) and is comparable to a $H$-$T$ phase diagram with both characteristic fields approaching zero at $T_C$.

To further characterize the charge transport, we present the Hall resistivity $\rho_{xy}$ in Fig. 6 for temperatures between 5 and 70 K for current applied in the $ab$-plane and magnetic fields oriented along the crystalline $c$-axis. Data for three representative temperatures are shown in Fig. 6 (a). Above $T_C$ (at 70 K in the figure), $\rho_{xy}$ is linear in $H$, consistent with hole-like carriers. We note that the Hall voltage we measure is substantially smaller than that in $Co_{1/3}NbS_2$ [37] and $Cr_{1/3}NbS_2$ [33], suggesting a higher carrier density. As $T$ decreases below 50 K, an anomalous Hall effect (AHE) becomes apparent in $\rho_{xy}$ with a strong negative contribution at low fields. The magnitude of the AHE is maximum in proximity to $T_C$, decreasing at lower temperatures as expected since $\rho_{ab}$ decreases substantially below the magnetic ordering temperature [Fig. 4 (a)]. The standard analysis of the Hall effect in magnetic materials considers two contributions, with $\rho_{xy} = R_0H + 4\pi\ MR_S$, where $R_0$ and $R_S$ are the ordinary and anomalous Hall coefficients, respectively. In Fig. 6 (b) $\rho_{xy}/H$ as a function of $M/H$ is displayed for temperatures spanning $T_C$, demonstrating the linear dependence expected from this standard form and allowing an accurate determination of both $R_0$ and $R_S$. The values of $R_0$ and $R_S$ determined from linear fits to these data are shown in Figure 6(c), with $R_0$ displaying a decrease of more than a factor of 2 below $T_C$, suggesting a significant change to the electronic structure with magnetic ordering. Within a simple single band model, the estimated carrier concentration ($n = 1/|eR_0|$, where $e$ is the electronic charge) is $2.7\times 10^{21}$ cm$^{-3}$ at 70 K. In addition, $R_S$ approaches zero at 5 K consistent with the small low temperature $\rho_{ab}$, while the magnitude of $R_S$ near $T_C$ is large being ~10 times larger than that in MnSi near its $T_C$ of 29 K [38, 39]. We observe no contributions to $\rho_{xy}/H$ above background that are non-linear in $M/H$, which would indicate a topological Hall effect or the unusual behavior observed in $Cr_{1/3}NbS_2$ [35]. Figure 6(d) presents the variation of the anomalous

Hall parameter $S_H = R_S/\rho_{ab}^2$ as a function of $T/T_C$ with no apparent $T$-dependence, suggesting that the anomalous Hall effect is intrinsic in character.

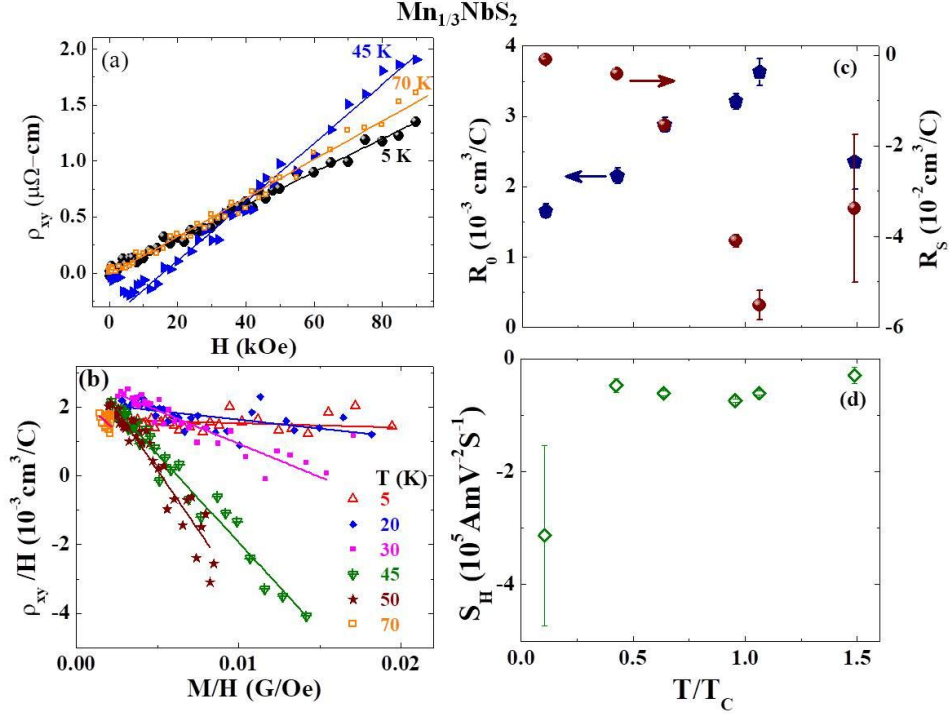

Fig. 6. Hall Effect for $Mn_{1/3}NbS_2$. (a) Hall resistivity, $\rho_{xy}$, vs. magnetic field, $H$, at 5, 45, and 70 K, (b) $\rho_{xy}/H$, vs. the magnetization divided by field ($M/H$), (c) Ordinary Hall coefficient $R_0$ and anomalous Hall coefficient, $R_S$, vs. temperature divided by the Curie temperature ($T/T_C$), (d) Anomalous Hall parameter, $S_H$ ($= R_S/\rho_{ab}^2$) vs. $T/T_C$.

The specific heat capacity divided by $T$, $C_P(T)/T$, of $Mn_{1/3}NbS_2$ as a function of $T$ is displayed in Fig. 7a along with that of $Cr_{1/3}NbS_2$ from Ghimire *et al.* [33]. A sharp asymmetric peak is observed at the onset of magnetic ordering, indicating a second order phase transition, similar in magnitude to that observed in $Cr_{1/3}NbS_2$. However, the magnitude of $C_P(T)/T$ of $Mn_{1/3}NbS_2$ remains much larger than that of $Cr_{1/3}NbS_2$ for all $T<T_C$, and the two display very different temperature dependencies. That the extra contribution to $C_P(T)/T$ in $Mn_{1/3}NbS_2$ originates from magnetic contributions is made clear by the large changes observed with the application of magnetic fields, shown in Fig. 7b. In addition, whereas $C_P/T$ of $Cr_{1/3}NbS_2$ at low temperature is well fit by a standard form for metals, $C_P/T=\gamma + \beta T^2$, with $\gamma$ proportional to the density of states of the charge carriers and β related to Debye temperature, that of $Mn_{1/3}NbS_2$ cannot be well described by this form due to the significant magnetic contributions that survive to the lowest temperature measured (1.5 K). Including a term commonly used to model the $C_P/T$ contribution due to magnons in ferromagnets, $\delta T^{1/2}$, did not result in reasonable values for the fit parameters. However, the similarity of $C_P/T$ for $Mn_{1/3}NbS_2$ and $Cr_{1/3}NbS_2$ above $T_C$ suggests a similar Debye temperature. The large low temperature contribution to $C_P/T$ for $Mn_{1/3}NbS_2$ implies that either $Mn_{1/3}NbS_2$ has a very small magnon stiffness, or there is a large density of other magnetic excitations available at low energies.

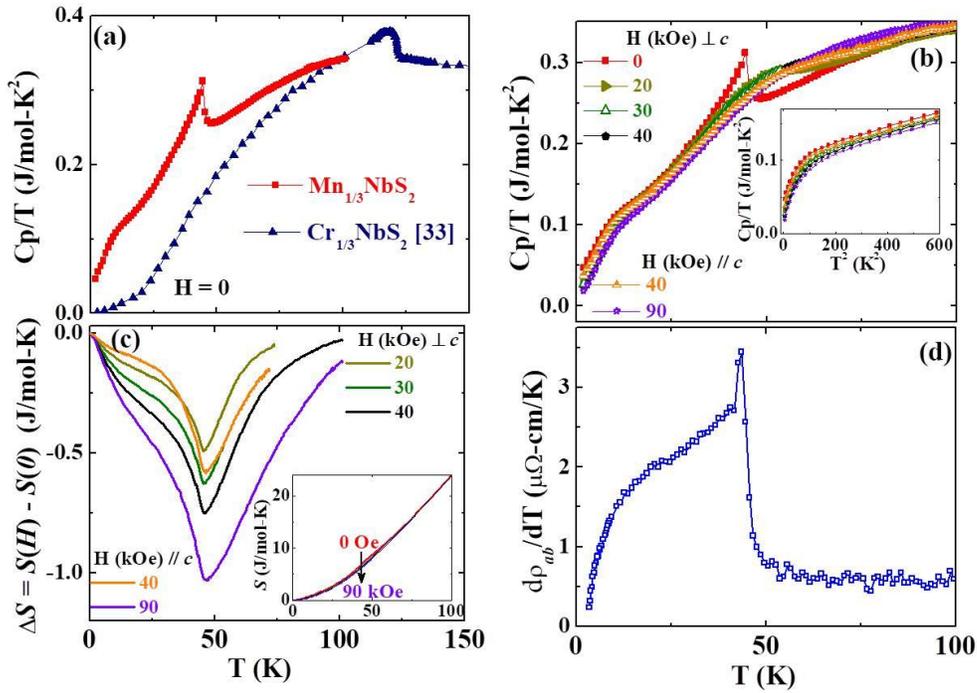

Fig. 7. Temperature and field dependence of the specific heat capacity of $Mn_{1/3}NbS_2$. (a) Specific heat capacity at constant pressure, $C_P$, divided by the temperature, $T$, vs. $T$ at zero magnetic field for $Mn_{1/3}NbS_2$. Data for $Cr_{1/3}NbS_2$ taken from Ref. [33] with permission of the authors. (b) $T$ dependence of $C_P$ for $0 \leq H \leq 9$ T. Inset: $C_p/T$ vs. $T^2$ at the same fields as in the main frame. (c) Change in entropy with application of field, $\Delta S = S(H) - S(0)$, vs. $T$. Inset: The magnetic entropy, $S$, determined from the integral of the data shown in frame (b). (d) The $T$ derivative of the resistivity, $d\rho_{ab}/dT$ vs. $T$.

The effect of a magnetic field on the $C_P(T)/T$ is displayed over a wide temperature range in Fig. 7b for fields parallel or perpendicular to the *c*-axis. Fields of a few tens of kOe are observed to effectively move entropy associated with the magnetic phase transition at $T_C$ to temperatures of up to 80 K. This suggests large entropy changes with moderate magnetic fields. There are also significant differences in $C_P(T)/T$ with the orientation of the magnetic field particularly at low temperature with a stronger field dependence seen for fields perpendicular to the *c*-axis.

To quantify the entropy, *S*, variation with field and temperature, *S(T,H)* was found by numerically integrating $C_P(T)/T$ after extrapolation to *T*=0, $S = \int_0^T \frac{C_p}{T} dT$. *S(T,H)* calculated in this manner is displayed for a wide range of *T* and *H* shown in the inset to Fig. 7c. The changes in *S* are further highlighted in the main frame of Fig. 7c, where $\Delta S = S(H) - S(0)$ is presented and where significant changes are apparent across the entire temperature range probed for all fields. As expected, the largest $\Delta S(H)$ occurs near $T_C$. However, there are large decreases in entropy apparent over a wide temperature range, including for temperature $T \ll T_C$ supporting the identification of the low temperature contributions to $C_P(T)/T$ as originating from magnetic excitations. Furthermore, the anisotropy noted in the discussion of $C_P(T)/T$ is more obvious in Fig. 7c with the field orientation dependence of $\Delta S$ reflecting the smaller saturation fields for fields oriented perpendicular to the *c*-axis. The similarity of the temperature derivative of the resistivity, $d\rho_{ab}/dT$, to $C_P(T)/T$ is apparent in Fig. 7d makes clear that the availability of a large density of low energy magnetic excitations contributes significantly to the scattering of carriers. This observation strongly supports our assertion that the MR results from the suppression of magnetic scattering with field.

## IV. Discussion and Conclusions

The previous section presented measurements that give an initial impression of the magnetic structure, the physical properties effected by the magnetic ordering, and the effect of magnetic excitations on the specific heat capacity and charge transport both at $H = 0$ and for moderately sized magnetic fields for $Mn_{1/3}NbS_2$. The motivation was provided by the discovery of a magnetic soliton lattice in isostructural $Cr_{1/3}NbS_2$ in small fields aligned parallel to the $NbS_2$ planes [9]. The data presented here reveal many similarities between the two compounds along with indications of a smaller *q* for helimagnetic ordering and a more complex behavior in the charge transport, *ac* susceptibility, and the magnetic contributions to the specific heat capacity for $Mn_{1/3}NbS_2$. Perhaps the most significant of these is the specific heat capacity, which reveals the presence of a much larger population of lower lying excited states that are greatly affected by magnetic fields. These same excitations are likely to be the main scattering mechanism for charge carriers revealed by the MR, as well as the structure observed in the *ac* susceptibility between 10 and 40 kOe. The conclusion that follows directly from these observations is that the magnetic ordering and its response to magnetic fields is far richer in $Mn_{1/3}NbS_2$ than in $Cr_{1/3}NbS_2$, despite the crystalline structural similarity. This is important because the chiral crystal structure of these materials strongly suggests that magnetic excitations are topologically nontrivial and may offer the same advantages, including a robustness against decay to a topologically trivial magnetic state, that have attracted enormous attention to the skyrmion-lattice-hosting materials [1-5].

One possible source of low-lying states that may not offer topological excitations is the Mn site disorder that we have identified in our single crystal x-ray diffraction analysis. Here, we found that ~15% of the intercalated Mn atoms reside on the crystallographic *2b* site of the hexagonal lattice. The disorder introduced into the magnetic ordering and excitation spectra associated with this population of magnetic moments is not yet known. In addition, a comparison to the $Cr_{1/3}NbS_2$ samples used in previous measurements that found such interesting magnetic structures is not possible, as the level of site disorder for Cr is not reported. This is also true for published measurements on $Mn_{1/3}NbS_2$. However, our estimates of the entropy contribution expected for this population of magnetic moments based upon the measured magnetic moments that we have deduced from our magnetization measurements, and the density of defect sites indicated by our X-ray data is $S_{mag} = nR \ln(2J+1) = 0.44$ J/mol K. This can account for less than $1/7^{th}$ of the entropy difference between $Mn_{1/3}NbS_2$ and $Cr_{1/3}NbS_2$ apparent in the inset to Fig. 7c ($S_{mag} = 3.25$ J/mol K below 37 K), even when accounting for the differences in magnetic ordering temperatures. Therefore, we conclude that the site disorder alone is not the cause for the large differences between the $Mn_{1/3}NbS_2$ and $Cr_{1/3}NbS_2$, at least in the simple case where the Mn atoms residing on the *2b* sites act as relatively uncoupled magnetic moments. Thus, the enhanced specific heat capacity at temperatures well below $T_C$ must involve the Mn on the *2c* site, which orders at 45 K, implying that the constraints on the magnetic structure due to the crystal structure symmetry must also hold for these low-lying modes. A second possible source for low-lying magnetic excitations could follow from stacking disorder of the $NbS_2$ planes along the *c*-axis, although more in-depth investigation of the crystal structure of our samples is needed to assess this possibility.

In summary, we have measured the structural, magnetic, magnetic structural, charge transport, and thermodynamic properties of $Mn_{1/3}NbS_2$ and compared them with published data on $Cr_{1/3}NbS_2$. We find largely similar behavior, a nearly ferromagnetic ordering along with a longer length scale modulation, and a resistivity and Hall effect consistent with a strong coupling of magnetic and charge transport properties. In addition, the specific heat capacity indicates a large density of low-lying magnetic excitations that is absent in $Cr_{1/3}NbS_2$. The low temperature specific heat capacity is broadly consistent with a very low spin wave stiffness, but is more likely to be associated with topologically interesting magnetic modes. The appearance of a positive MR for *H // c* that correlates well to features in the magnetic response and the magnetic contribution to the specific heat capacity, as well as a decrease in the width of the magnetic scattering peaks for neutron diffraction, may be indicators of these topological excitations. Thus, further investigation of the magnetic structure and excitation spectrum of the material will likely yield fruitful discoveries in this prototypical chiral structured magnet.

**Acknowledgements**

This material is based upon the work supported by the U.S. Department of Energy under EPSCoR Grant No. DE-SC0012432 with additional support from the Louisiana Board of Regents. PWA acknowledge the financial support of the US Department of Energy, Office of Science, Basic Energy Sciences, under Award No. DE-FG02-07ER46420. A portion of this work was performed at the National High Magnetic Field Laboratory, which is supported by the National Science Foundation Cooperative Agreement No. DMR-1644779 and the State of Florida. The work at ORNL's HFIR was sponsored by the Scientific User Facilities Division, Office of Science, Basic Energy Sciences (BES), U.S. Department of Energy (DOE). H. B. C. acknowledges support of U.S. DOE BES Early Career Award No. KC0402010 under Contract No. DE-AC05-00OR22725

Supplemental Materials for Observation of a mesoscopic magnetic modulation in chiral Mn$_{1/3}$NbS$_2$

Sunil K. Karna[1], F. N. Womack[1], R. Chapai[1], D. P. Young[1], M. Marshall[2], Weiwei Xie[2], D. Graf[3], Yan Wu[4], Huibo Cao[4], L. DeBeer-Schmitt[4], P. W. Adams[1], R. Jin[1], and J. F. DiTusa[1]

[1]Department of Physics and Astronomy, Louisiana State University, Baton Rouge, Louisiana 70803, USA
[2]Department of Chemistry, Louisiana State University, Baton Rouge, Louisiana 70803, USA
[3]National High Magnetic Field Laboratory, Florida State University, Tallahassee, Florida 32310, USA
[4]Neutron Scattering Division, Oak Ridge National Laboratory, Oak Ridge, TN 37831, USA

Table 1-3 shows the results from the single crystal x-ray diffraction refinement which are discussed in the manuscript. Table 1 lists the lattice parameters, goodness of fit and the number of reflections used in single crystal x-ray diffraction refinement for **Mn$_{1/3}$NbS$_2$** crystals. Table 2 includes the atomic coordinates obtained from the single crystal x-ray diffraction refinement while Table 3 presents the thermal displacement parameters.

**Table 1.** Single crystal refinement for Mn$_{1/3}$NbS$_2$ at 296 (2) K.

| Refined Formula | Mn$_{0.85(2)}$Nb$_3$S$_6$ |
|---|---|
| F.W. (g/mol) | 257.93 |
| Space group; Z | $P6_322$; 4 |
| $a$(Å) | 5.783 (5) |
| $c$(Å) | 12.644 (11) |
| V (Å$^3$) | 366.3 (7) |
| Extinction Coefficient | 0.047 (7) |
| θ range (deg) | 4.069-33.078 |
| No. reflections; $R_{int}$ | 3495; 0.0733 |
| No. independent reflections | 475 |
| No. parameters | 20 |
| $R_1$: $\omega R_2$ ($I>2\sigma(I)$) | 0.0448; 0.1344 |
| Goodness of fit | 1.455 |
| Diffraction peak and hole (e$^-$/ Å$^3$) | 6.178; -1.481 |

**Table 2.** Atomic coordinates and equivalent isotropic displacement parameters of $Mn_{1/3}NbS_2$ system. ($U_{eq}$ is defined as one-third of the trace of the orthogonalized $U_{ij}$ tensor ($Å^2$))

| Atom | Wyck. | Occ. | x | y | z | $U_{eq}$ |
|------|-------|------|---|---|---|------|
| Nb1 | 2a | 1 | 0 | 0 | 0 | 0.0043(5) |
| Nb2 | 4f | 1 | 1/3 | 2/3 | 0.5006(1) | 0.0045(5) |
| Mn3 | 2c | 0.85(2) | 1/3 | 2/3 | ¼ | 0.0055(5) |
| S4 | 12i | 1 | 0.3330(2) | 0.0035(2) | 0.1239(2) | 0.0058(6) |

**Table S3.** Anisotropic thermal displacements from $Mn_{1/3}NbS_2$

| Atom | U11 | U22 | U33 | U23 | U13 | U12 |
|------|-----|-----|-----|-----|-----|-----|
| Nb1 | 0.0007 (6) | 0.0007 (6) | 0.0116 (8) | 0 | 0 | 0.0003 (3) |
| Nb2 | 0.0009 (6) | 0.0009 (6) | 0.0118 (7) | 0 | 0 | 0.0004 (3) |
| Mn3 | 0.0033 (5) | 0.0033 (5) | 0.0097 (7) | 0 | 0 | 0.0017 (3) |
| S4 | 0.0022 (7) | 0.0023 (7) | 0.0129 (9) | -0.0001 (3) | -0.0005 (3) | 0.0012 (4) |